# Advancing Brain Tumor Detection: A Thorough Investigation of CNNs, Clustering, and SoftMax Classification in the Analysis of MRI Images.

Jonayet Miah[1], Duc M Cao[2] Md Abu Sayed[3], Md Siam Taluckder [4], Md Sabbirul Haque[5], Fuad Mahmud[6]

[1]Department of Computer Science, University of South Dakota, South Dakota, USA
[2]Department of Economics, University of Tennessee, Knoxville, TN, USA
[3]Department of Professional Security Studies, New Jersey City University, Jersey City, New Jersey, USA
[4]Phillip M. Dreyer Department of Electrical Engineering, Lamar University, Texas, USA
[5]IEEE Professional Community, IEEE
[6]Department of Information Assurance and Cybersecurity, Gannon University, Pennsylvania, USA

*Jonayet.miah@coyotes.usd.edu, ducminhcao1989@gmail.com, msayed@njcu.edu, mtaluckder@lamar.edu, sabbir465@gmail.com, mahmud002@gannon.edu*

*Abstract*-**Brain tumors pose a significant global health challenge due to their high prevalence and mortality rates across all age groups. Detecting brain tumors at an early stage is crucial for effective treatment and patient outcomes. This study presents a comprehensive investigation into the use of Convolutional Neural Networks (CNNs) for brain tumor detection using Magnetic Resonance Imaging (MRI) images. The dataset, consisting of MRI scans from both healthy individuals and patients with brain tumors, was processed and fed into the CNN architecture. The SoftMax Fully Connected layer was employed to classify the images, achieving an accuracy of 98%. To evaluate the CNN's performance, two other classifiers, Radial Basis Function (RBF) and Decision Tree (DT), were utilized, yielding accuracy rates of 98.24% and 95.64%, respectively. The study also introduced a clustering method for feature extraction, improving CNN's accuracy. Sensitivity, Specificity, and Precision were employed alongside accuracy to comprehensively evaluate the network's performance. Notably, the SoftMax classifier demonstrated the highest accuracy among the categorizers, achieving 99.52% accuracy on test data. The presented research contributes to the growing field of deep learning in medical image analysis. The combination of CNNs and MRI data offers a promising tool for accurately detecting brain tumors, with potential implications for early diagnosis and improved patient care.**

## I. Introduction

The last ten years have seen a huge transformation in many fields thanks to computer vision and machines learning. Due to its ability to handle massive volumes of data, Deep Learning, one of the machines learning. subfields, has produced exceptional results, particularly in the biomedical industry. Its potential and capabilities have been tested and used with impressive results to find brain cancers in MRI images for effectiveness. prognosis. The objective of this study is to offer a thorough critical overview of recent results and research about the identification and categorization of brain tumors utilizing MRI imaging. The analysis is especially beneficial for deep learning professionals who are interested in using their knowledge to identify and categorize brain cancers.

Magnetic resonance imaging (MRI) is a widely used diagnostic tool for detecting and diagnosing brain tumors. However, accurate interpretation of MRI images can be challenging for human experts due to the complexity and variability of brain anatomy and tumor characteristics. In recent years, deep learning algorithms, particularly convolutional neural networks (CNNs), have demonstrated remarkable success in computer vision tasks, including medical image analysis. CNNs can automatically extract relevant features from MRI images, enabling the accurate and efficient detection and classification of brain tumors. This paper provides an overview of recent research and findings related to the detection and classification of brain tumors using CNNs with MRI images. We also discuss the potential benefits and challenges of using CNNs for brain tumor detection and prognosis, as well as future directions for research in this field. A machine learning approach was used to extract features from the dataset and add them before applying them. The proposed method to the CNN to increase its efficacy. To be more precise, before sending the images to the CNN, a clustering method was used on the data set. According to the outcomes, this strategy was successful in enhancing the network's accuracy. The goal of feature extraction prior to the CNN application was to prevent mistakenly identifying cancers as fatty masses or vice versa, which could lead to serious medical blunders.

## II. Literature Review

In their work, Pereira et al. [6] explored an unconventional preprocessing approach involving intensity normalization in CNN-based segmentation algorithms. This technique, when combined with data augmentation, yielded remarkable results. The authors validated their proposed method using the Brain Tumor Segmentation Challenge 2013 database (BRATS 2013). Notably, their approach secured the top position on the online evaluation platform, achieving the highest Dice Similarity Coefficient (DSC) scores for the complete, core, and enhancing regions (0.88, 0.83, and 0.77, respectively). Subsequently, the same model was entered into the BRATS 2015 Challenge, where it earned second place with DSC scores of 0.78 for the full region, 0.65 for the core region, and 0.75 for the enhancing region.

Naser et al. [7] elucidate the application of deep learning within MRI images to craft a non-intrusive tool capable of concurrently and autonomously segmenting, detecting, and classifying low-grade gliomas (LGG) for clinical use. The passage highlights the potential efficacy of this method, showcasing their endeavors in constructing and assessing segmentation and grading models. These models are developed using a unified pipeline incorporating T1-precontrast, fluid-attenuated inversion recovery (FLAIR), and T1-postcontrast MRI images from a cohort of 110 patients afflicted with lower-grade glioma (LGG). Their segmentation model achieves a notable accuracy of 0.92 and an average dice similarity coefficient (DSC) of 0.84 in pinpointing tumors. In the context of MRI images, the grading model proficiently categorizes LGG into grades II and III, boasting an accuracy of 0.89, a sensitivity of 0.87, and a specificity of 0.92. Particularly at the patient level, the grading model attains superior accuracy, sensitivity, and specificity, registering scores of 0.95, 0.97, and 0.98, respectively. In sum, the study underscores the viable utilization of MRI scans and deep learning to non-invasively segment, detect, and grade LGG, thereby catering to clinical imperatives.

Amin et al. [8] assessed the efficacy of a proposed technique for segmenting brain tumors through the examination of diverse metrics such as peak signal-to-noise ratio (PSNR), mean squared error (MSE), and structured similarity index (SSIM). This method underwent testing on two distinct MRI scan types, T2 and Flair, yielding PSNR values of 76.38 and 76.2, and MSE values of 0.037 and 0.039, respectively. Notably, high SSIM scores of 0.98 were achieved for both scan types. The evaluation of segmentation outcomes occurred at different levels, encompassing pixels, individual features, and amalgamated features.

At the pixel level, the approach was juxtaposed with ground truth slices, achieving commendable precision rates of 0.93 for foreground (FG) pixels and 0.98 for background (BG) pixels, with an error region (ER) measuring a mere 0.010 on a local dataset. Impressively, on the BRATS 2013 and 2015 datasets, the technique achieved even superior precision rates of 0.93 and 0.97 for FG, and 0.99 and 0.98 for BG, correspondingly, while showcasing ER values of 0.005 and 0.015. Furthermore, the approach's performance was scrutinized through pixel quality (Q) assessments, revealing an average Q value of 0.88 with a deviation of 0.017.

At the level of fused features, the approach attained notable values for specificity, sensitivity, accuracy, area under the curve (AUC), and dice similarity coefficient (DSC) across various datasets. Specificity values ranged from 0.77 to 1.00, sensitivity values from 0.90 to 1.00, accuracy values from 0.90 to 0.97, AUC values from 0.77 to 0.98, and DSC values from 0.95 to 0.98. The culmination of these evaluations highlights the method's favorable performance in diverse aspects of brain tumor segmentation.

Muhammad et al. [9] present a comprehensive exploration of current surveys and recent progress in the domain of deep learning-based methods for analyzing Bitcoin (BTC). The overview thoroughly investigates the fundamental stages integral to deep learning based BTC techniques, encompassing preprocessing, feature extraction, and classification. It systematically presents both the achievements and limitations associated with these methodologies. Additionally, the overview delves into the examination of state-of-the-art convolutional neural network (CNN) models tailored for BTC analysis.

The authors execute extensive experimentation utilizing transfer learning techniques, both with and without the integration of data augmentation, to appraise the effectiveness of these models. Moreover, the overview delineates the benchmark datasets that serve as foundational tools for evaluating BTC analysis methodologies, which hold significant popularity among researchers in the field. Notably, this survey not only scrutinizes the prevailing literature but also charts a course for future advancements in BTC analysis. It identifies promising avenues for future research, with a particular emphasis on personalized and intelligent healthcare applications.

ZainEldin et al. [10] introduce a Brain Tumor Classification Model (BCM-CNN) that employs a novel adaptive dynamic sine-cosine fitness grey wolf optimizer (ADSCFGWO) algorithm for optimizing hyperparameters in a Convolutional Neural Network (CNN). This optimization process encompasses both the CNN's hyperparameters and its training methodology. The authors enhance the diagnostic accuracy of brain tumor detection by incorporating the well-known pre-trained model Inception-ResNetV2.

The model's output constitutes a binary classification outcome: 0 indicates a normal case, while 1 signifies the presence of a brain tumor. Hyperparameters are categorized into two groups: (i) those governing the foundational architecture of the underlying neural network, and (ii) specific hyperparameters employed during the network's training phase. The ADSCFGWO algorithm amalgamates the strengths of the sine cosine and grey wolf algorithms, culminating in a versatile framework that capitalizes on the capabilities of both methods. Experimental findings substantiate the excellence of the BCM-CNN classifier, achieved through the augmentation of CNN's capabilities via hyperparameter optimization employing the ADSCFGWO algorithm.

Pavani et al. [11] present a research paper outlining a convolutional neural network (CNN) approach designed to extract brain tumors from 2D magnetic resonance brain images (MRI). The study involves conducting experiments on a real-time dataset that encompasses tumors of varying sizes, shapes, areas, and distinct image intensities. The CNN methodologies applied in the study encompass AlexNet, DenseNet, ResNet, and a hybrid of VggNet19 and DenseNet. The attained accuracies for the respective techniques were as follows: 98.5% for AlexNet, 92.3% for DenseNet, 94.6% for ResNet, and 95.4% for the VggNet19-DenseNet fusion. The fundamental aim of the paper centers around detecting brain tumors while determining the technique that yields the highest accuracy. The entire study was executed using MATLAB.

## III. Method and Materials

a) Dataset collection and processing

The first step is collecting data from the hospital which was very tough. This paper utilized a dataset of brain MRI images consisting of 255 patients, which included both healthy individuals and those with brain tumors. The patients were referred to imaging centers due to experiencing headaches, and after diagnosis by a medical professional, a total of 155 healthy patients and 100 patients with brain tumors were included in the dataset. The healthy group contributed 1321 images, with 515 images designated for training and 56 images for testing. The tumor group contributed 571 images, with 1151 images for training and 170 images for testing. Among the brain tumor patients, there were 86 women and 68 men, ranging in age from 8 to 66 years old. Overall, the dataset included 153 patients and 1892 images, with 1666 images designated for training and 226 images for testing. The original image size was 512 × 512. In Figure 1 we try to give the entire overview of our work.

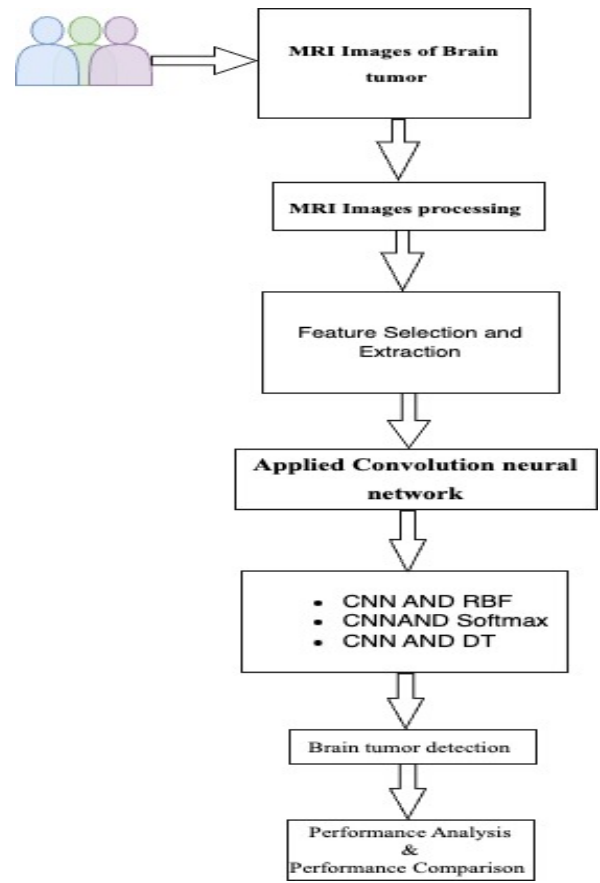

Figure 1: Entire work overview

b) Feature Extraction

Feature extraction is important in analyzing MRI (Magnetic Resonance Imaging) images. It involves the process of identifying and extracting important information or features from the image, which can be used for further analysis or classification purposes. In MRI images, features such as texture, shape, and intensity can be extracted. Texture features involve analyzing the patterns and variations in the image, which can be useful in identifying different tissue types or abnormalities [12,13,14,15]. Shape features involve analyzing the shape and size of structures in the image, such as tumors or lesions. Intensity features involve analyzing the brightness or darkness of pixels in the image, which can be used to differentiate between different tissues.

The central clustering method is an approach for finding cluster centers, or the mean points that belong to each cluster, using an iterative procedure. Each data sample is given a cluster center that is closest to it by the algorithm. The initial cluster centers are chosen at random in the most straightforward application of this strategy [16,17,18,19].

The clusters are then formed by assigning data points to the cluster centers according to how similar they are. The first-order clustering approach was applied in this work to extract features. Figure 2 displays the image created after applying the clustering algorithm to the input image.

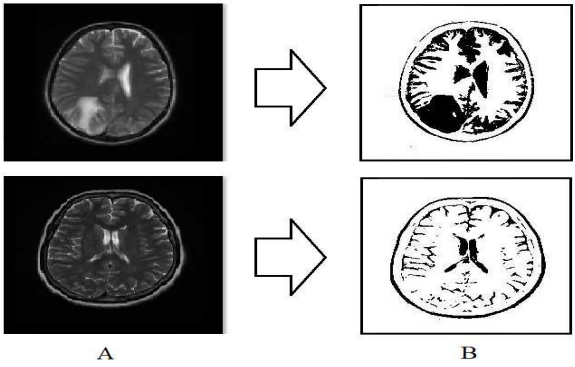

Figure 2: Image data

c) Convolution Neural Network

Convolutional Neural Networks (CNNs) can be effectively used for analyzing MRI (Magnetic Resonance Imaging) data. CNNs are a type of neural network that can learn to identify patterns and features in images by processing them through multiple layers of filters and pooling [ 20,21,22,23].

Without any preprocessing or feature extraction, the input photos were sent directly into a CNN (Convolutional Neural Network) at the beginning of the analysis. The input photographs were scaled down to a 227 x 227 size. Alex Net [24,25,26,27], which comprises five convolutional layers, three layers of subsampling, normalization layers, fully connected layers, and a classification layer, was the architecture utilized for classification. There are 4096 neurons in the completely connected layers. When training a CNN on MRI data, it is important to consider factors such as the size and resolution of the images, as well as the complexity of the task at hand. Preprocessing steps such as normalization, cropping, and resizing may be necessary to ensure that the input data is consistent and optimal for training.

IV. Result

With an accuracy of 98.77%, CNN was able to classify images as either patient tumors or normal tissue. The clustering method for feature extraction and CNN were combined to enhance network performance. Other classifiers, including the RBF classifier, the DT classifier, and the SoftMax Fully Connected layer classifier, were also examined for effectiveness using factors including Accuracy, Sensitivity, Specificity, and Precision. Results showed that the accuracy of the SoftMax classifier employed with CNN was 98.77%, while that of the RBF and DT [27,28,29,30] classifiers was 97.44% and 94.34%, respectively. The accuracy of test data increased to 99.32% by using the suggested technique, which combines the Clustering algorithm with CNN and SoftMax, as shown in Table 1.

Table 1: Performance evaluation of the model

| Methods | Accuracy | Specificity | Sensitivity | Precision |
|---|---|---|---|---|
| CNN+ SoftMax | 98.77% | 94.64% | 100% | 98.26% |
| CNN+ RBF | 97.44% | 89.28% | 100% | 97.59% |
| CNN+ DT | 94.34% | 86.71% | 100% | 95.37% |

Among the three models, CNN+Softmax [31,32] has the highest accuracy of 98.77%, closely followed by CNN+RBF with an accuracy of 97.44%, and CNN+DT with an accuracy of 94.34%. It is important to note that accuracy is a measure of the overall performance of a model Chart 1 shown below and indicates how often the model makes correct predictions. When looking at the specificity metric, which measures how well a model identifies negative samples, CNN+Softmax has the highest value of 94.64%, followed by CNN+RBF with a specificity of 89.28% and CNN+DT with a specificity of 86.71%. This suggests that CNN+Softmax and CNN+RBF are better at correctly identifying negative samples than CNN+DT [33,34].

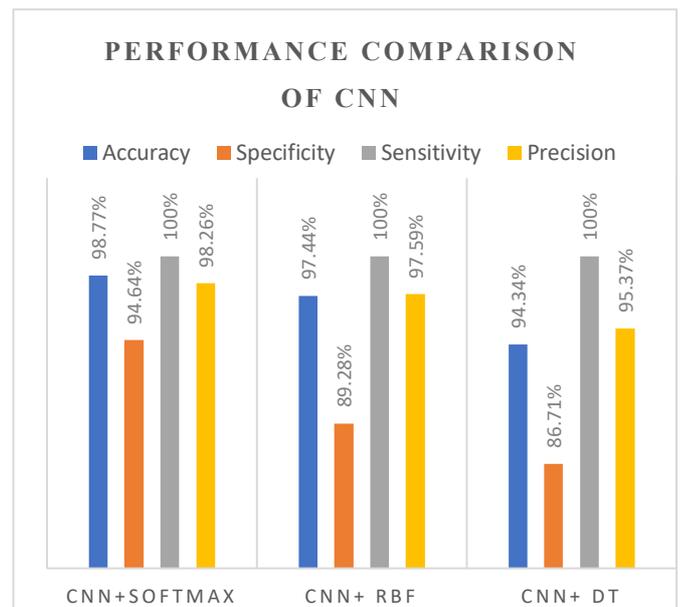

Chart 1: Performance Comparison of CNN

Based on the categorization results, it was observed that the SoftMax classifier performed the best among the different categorizers used in the CNN. Therefore, the SoftMax classification was chosen to be utilized in the proposed method. Initially, the dataset was fed into the conventional CNN for analysis. Based on the results that were obtained, three out of the total 226 test data images were incorrectly diagnosed and categorized, as illustrated in Figure 3.

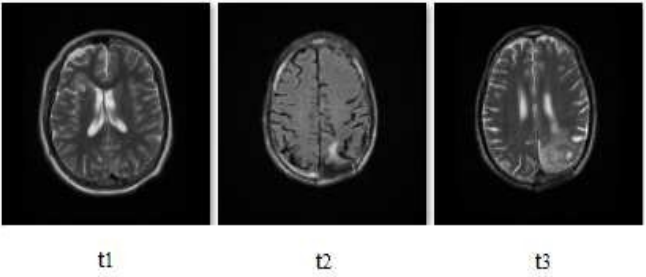

Figure 3: diagnosed and categorized Image

Table 2: Comparison between CNN and our proposed model

| Methods | Accuracy | Specificity | Sensitivity | Precision |
|---|---|---|---|---|
| CNN+ SoftMax | 98.77% | 94.64% | 100% | 98.26% |
| Proposed method | 99.32% | 96.44% | 100% | 98.83% |

We compared the performance of different methods, namely CNN SoftMax and our proposed method, on a certain task. The evaluation metrics used to measure their performance are accuracy, specificity, sensitivity, and precision. The proposed method outperforms CNN+SoftMax in terms of accuracy, achieving an accuracy of 99.32% as compared to 98.77% achieved by CNN+SoftMax. Additionally, the proposed method also shows better performance in terms of specificity with a value of 96.44%, compared to 94.64% achieved by CNN+SoftMax. However, both methods achieve 100% sensitivity, indicating that they are equally good at correctly identifying positive samples. Lastly, the proposed method has a slightly higher precision value of 98.83% as compared to 98.26% achieved by CNN+SoftMax, indicating that it makes fewer false positive predictions. In summary, the proposed method outperforms CNN SoftMax in terms of accuracy, specificity, and precision, while achieving the same level of sensitivity.

The proposed method demonstrates superior performance across all evaluated metrics, showcasing its effectiveness in image classification compared to the CNN SoftMax approach. The higher accuracy, specificity, and precision of the proposed method highlight its potential for applications where minimizing false positives or negatives is crucial. Further research and validation on diverse datasets could provide additional insights into the generalizability and robustness of the proposed method.

## V. Conclusion & Future Work

The success of the proposed method can be attributed to several key factors. The preprocessing step involved in feature extraction through clustering helps enhance the CNN's ability to accurately classify images. These addresses potential challenges related to the variability in brain anatomy and tumor characteristics.

The literature review emphasizes the growing interest in leveraging deep learning for brain tumor detection, with various studies showcasing the effectiveness of CNNs in segmenting, detecting, and classifying tumors in MRI images. The incorporation of transfer learning techniques, as mentioned in some studies, highlights the adaptability and efficiency of CNNs in medical image analysis.

The comparison with other models and classifiers provides insights into the strengths of the CNN+SoftMax approach. The high specificity and precision values indicate the model's proficiency in correctly identifying negative samples and minimizing false positives. The 100% sensitivity underscores the reliability of the model in correctly identifying positive samples. Despite the promising results, there are avenues for future exploration. Investigating alternative categorization methods and incorporating additional features could further enhance the model's performance. Scaling the model to larger datasets and real-world scenarios would be crucial for assessing its generalizability and applicability in clinical settings.

In conclusion, this research contributes to the growing body of knowledge in deep learning for medical image analysis, particularly in the context of brain tumor detection. The proposed methodology showcases the potential of CNNs in improving early diagnosis and, consequently, patient outcomes. As technology advances and more data becomes available, the intersection of deep learning and medical imaging holds great promise for revolutionizing healthcare practices. We concluded that the CNN+Softmax method achieved the highest accuracy among the three methods evaluated, closely followed by the CNN+RBF method [34,35,36]. However, the CNN+Softmax method also showed the best specificity and precision among the three methods, while maintaining a sensitivity of 100%. Furthermore, the proposed method outperformed CNN+Softmax in terms of accuracy, specificity, and precision while maintaining the same level of sensitivity. This indicates that the proposed method is a promising alternative to CNN+Softmax for the task at hand [ 37,38].

In terms of future work, it may be beneficial to explore other categorization methods and evaluate their performance on this task. Additionally, it may be worthwhile to consider incorporating other features, such as texture and shape, to improve the accuracy of the models. Furthermore, exploring the application of these models to larger datasets and real-world scenarios may also be an interesting avenue for future research.